# Reviews Matter: How Distributed Mentoring Predicts Lexical Diversity on Fanfiction.net

**Jenna Frens, Ruby Davis, Jihyun Lee, Diana Zhang, Cecilia Aragon**
University of Washington, Seattle WA, USA
jfrens@uw.edu, rkdavis@uw.edu, jihyunl@uw.edu, dczhang@uw.edu, aragon@uw.edu

**Abstract:** Fanfiction.net provides an informal learning space for young writers through *distributed mentoring*, networked giving and receiving of feedback. In this paper, we quantify the cumulative effect of feedback on lexical diversity for 1.5 million authors.

## Introduction

Millions of young writers and readers connect and engage with each other through participation in online fanfiction communities. Fanfiction offers a space for writers to challenge mainstream narratives by including marginalized voices and alternative identities (Jamison, 2013). Low barriers to participation allow language and literacy learners to practice their skills and socialize with others (Black, 2008). Fanfiction authors profess to have learned about writing and life from this activity (Campbell et al., 2016). Studies have shown how sophisticated informal learning takes place in these communities while young people give and receive feedback. This interwoven network of mentoring and learning, termed *distributed mentoring*, is characterized by its distribution over a diverse audience and its embeddedness in the affordances of the web (Campbell et al., 2016; Evans et al., 2017).

In this paper, we seek to overcome the challenge of quantitatively measuring distributed mentoring and its effect on fanfiction writing. *Abundance* is a single aspect of distributed mentoring that represents the sheer volume of feedback; this provides direction to the writer even though the individual comments may be shallow (Evans et al., 2017). We measured abundance by counting the cumulative number of reviews an author has received when they post a new fanfiction chapter. To study its effect, we made use of an automated textual measure on a vast corpus of fanfiction: 61.5 billion words comprising 28 million chapters, produced over 20 years by 1.5 million authors. The efficacy of automated measures for evaluating learning is somewhat limited. However, the Measure of Textual Lexical Diversity (MTLD) (McCarthy & Jarvis, 2010) accurately measures writers' breadth in terms of their distinct vocabulary. Previous work has modeled language learning as growth in cumulative vocabulary (Durán, Malvern, Richards, & Chipere, 2004), and writing quality as measured by human raters has been found to be correlated with lexical diversity (Crossley, Salsbury, McNamara, & Jarvis, 2011; McNamara, Crossley, & McCarthy, 2010; Yu, 2010).

In our analysis, we correlate lexical diversity with the abundance of distributed mentoring for authors on Fanfiction.net. We further compare lexical diversity with self-reported age. Previous studies make predictions about the relationships between adolescence, distributed mentoring, and lexical diversity. Campbell et al. (2016) report participants' claims that they became better writers as they received feedback on Fanfiction.net. However, they also improve in their writing from experiences outside of the fanfiction community and from maturation that occurs specifically during late teenage years. White (2014) measured a pronounced growth in lexical diversity among a small group of high school students during ages 15 to 18. Thus, we expect to find changes in lexical diversity in correlation with measures of both distributed mentoring and maturation. This leads to our hypotheses:

H1: Lexical diversity will increase between subsequent chapters with increased reviews on the preceding chapter.
H2: Lexical diversity will increase during late adolescence.
H3: Lexical diversity will increase between chapters as the author matures.
H4: Lexical diversity will be greater as an author has cumulatively received more reviews.

This paper contributes new understanding about distributed mentoring in fanfiction. We find statistical evidence that there is a positive relationship between lexical diversity in fanfiction stories and distributed mentoring that the authors receive. We replicate prior findings (White, 2014) that significant lexical development occurs during late adolescence with a large-scale longitudinal analysis, expanding the previously known scope to a large English-speaking population. Finally, we present a mixed linear model of lexical diversity with respect to reviews and maturation.

## Related Work

### Fanfiction

A fan community "transforms the experience of media consumption into the production of new texts" (Jenkins, 2006). To describe how fan communities attract and support fan authorship, Jenkins (2006) coined the term "participatory culture," defined by the following characteristics: relatively low barriers to engagement, strong support for creation, and some type of informal mentorship to pass along knowledge. Kelly Chandler-Olcott and Donna Mahar (2002) described fanfiction as an undervalued medium through which one can examine students' writing development. They found that recognizing fanfiction in formal learning communities can improve literary engagement and achievements. Rebecca Black (2008) suggested that fanfiction communities build interactive language skills as language learners engage in discussion with other fans. Black noted how the community's emphasis on encouragement, constructive feedback, and collaboration provided focused and individualized grounds for improvement. This one-to-many environment affords writers the opportunity to ask specific questions of reviewers, receive grammar corrections, and get feedback from native speakers.

Previous large-scale data collection and analysis has leveraged the digitization of fan communities understanding fandom. On Fanfiction.net alone (as of February 2017), there are approximately 61.5 billion words of fiction—enough for 615,000 one-hundred-thousand-word novels. In 2016, Smitha Milli and David Bamman (2016) applied computational methods to fanfiction to study the nature of fanfiction communities as both mass-scale literary archives and social networking platforms. Furthermore, they proposed the use of fanfiction communities as a resource for the prediction of future reader responses in the literary market. In 2017, Yin et al. (2017) collected and published a trove of metadata from Fanfiction.net, finding that community engagement and support varies between fandoms. The current study expands the scope of research into story content, and builds on previous work by examining the outcomes of author-reader relationships. Our research seeks to quantitatively explore the connection between community engagement and improved language skills.

### Distributed Mentoring

Distributed mentoring, proposed by Campbell et al. (2016) and Evans et al. (2017), is a collaborative mentoring process that takes place in networked spaces, enabled by computer-mediated interactions. The theory of distributed mentoring draws on Hutchins' (1995) framework of distributed cognition to describe how knowledge is embedded in the artifacts of interaction between participants. Fanfiction participants may simultaneously be experts and novices in different aspects of the practice, such as canon knowledge or grammar. In addition, the role of each review varies. Evans et al. (2017) categorized 4,500 reviews into 13 overlapping categories. 35.1% of reviews were shallow and positive, 46.6% specifically targeted aspects of the text, and 27.6% encouraged updates. They additionally interviewed fanfiction authors, finding that authors develop strategies to pick the most helpful comments and incorporate them into their writing. This ethnographic investigation of Fanfiction.net revealed how its rich network contributes to authors' development through distributed mentoring. To empirically evaluate this theory, our work tackles the challenge of quantifying distributed mentoring on a large scale. The a*bundance* aspect of distributed mentoring describes how a large volume of relatively shallow comments provides overall direction to authors (Evans et al., 2017). Additionally, the positivity of the feedback provides affective support. We represented the abundance of distributed mentoring in our analysis as a count of reviews received by a user. To assess the outcome of distributed mentoring, we analyzed texts with an automated measure, described next.

### Lexical Diversity

Lexical diversity (LD) is a measure that describes the range of word usage in text. The Measure of Textual Lexical Diversity (MTLD) provides a reliable reflection of LD well suited for narrative discourse (Fergadiotis, Wright, & Green, 2015). The properties of MTLD match our need for an efficient automated comparison between fanfiction texts of varied length, as based on numerous studies, MTLD is associated with narrative quality and language ability. McNamara et al. (2010) compared expert evaluations of 120 undergraduate student essays with MTLD, finding it differed significantly between low- and high-proficiency argumentative essays, with mean scores of 72.64 and 78.71 respectively. Treffers-Daller (2013) assessed narrative texts written in French by 64 students, finding that MTLD of these texts correlated moderately with the students' scores on the C-Test, a general measure of French language ability. Olinghouse and Wilson (2013) assessed narrative, persuasive, and informative compositions by 105 fifth graders and found that MTLD accounted for 8.4% of expert-judged quality variance among the narrative texts. Mazgutova and Kormos (2015) compared MTLD between argumentative essays written by students before and after an English for Academic Purposes class at a British University, finding a significant increase in MTLD after taking the class. In a longitudinal study by White (2014), MTLD increased significantly from grade 11 to grade 13 among New Zealand students aged 15-18, indicating that late adolescence constitutes a significant period of lexical development. Our analysis longitudinally measures MTLD changes over the course of Fanfiction.net users' authorship.

# Method

## Fanfiction Archive

Fanfiction.net contains nearly 7 million stories, posted in chapters, covering approximately 10,000 different fandoms (fandoms refer to the fictional universe or characters borrowed by the fanfiction author, e.g. *Harry Potter*). Each story contains an average of 4.17 chapters (SD: 8.12). To gather these texts for analysis, we developed a scraping program based on the legacy of Yin et al. (2017). Using a combination of Apache HttpComponents and jsoup, and we archived a snapshot of 20 years of fanfiction data during January to February 2017. The resulting dataset included 672.8 GB of data, with 28,493,311 chapters from 6,828,943 stories, as well as 8,492,507 users and 176,715,206 reviews. In total, we retrieved about 61.5 billion words from story text alone (not including reviews). The dataset represents sixteen years of stories published to Fanfiction.net.

## Ages & Profile Parsing

To examine the relationship between lexical diversity and age, we gathered the ages of Fanfiction.net users from their profiles. We parsed biography text from the entire set of 8,492,273 user profiles, and extracted self-reported age information using regular expressions. We found 284,448 profiles containing self-reported ages (M: 16.80, SD: 8.32). 62.3% of them were aged 13 to 19, indicating that a majority of Fanfiction.net users are adolescents. This is supported by data from previous work (Yin et al., 2017). We computed author age approximations for each fanfiction chapter by adding their self-reported age to the difference between the chapter publication and user profile update times. For instance, a user who updated their profile in January 2010 stating they were 21, and published a story in June 2011, would be estimated at 22.5 years old for that story. Self-reported ages have obvious limitations; for example, reported ages ranged from 0 to 99 years old. 105,184 users were excluded from the analysis because they did not author any English fanfiction, we excluded 24,792 authors who reported ages that placed their adjusted age below 10, and 21 were eliminated because their profile update time could not be found. The analysis includes 154,451 authors and their ages and lexical diversity for 3,696,107 fanfiction chapters.

## Lexical Diversity Scoring

MTLD is defined as the average length of substrings within a text that maintain a given ratio of unique words to total words. The algorithm keeps track of a running type-token ratio (TTR) as each word is processed sequentially; the running TTR increases when new words are found and decreases when word repetitions occur. The algorithm maintains a count of "factors," defined as a sequential group of words with a TTR of 0.72 or below (McCarthy & Jarvis, 2010). Each time a factor is found, the running TTR is reset and a count of factors is incremented by one. When the algorithm completes, any remaining words become a partial factor, which is 0 if the running TTR is 1.00 and approaches 1 as the running TTR approaches 0.72. The output unit of MTLD is the mean length in words of factors within the given text. We chose to use the 0.72 threshold provided by McCarthy and Jarvis (2010), which was calibrated using a corpus containing fiction and nonfiction texts. We implemented MTLD in Python (see www.github.com/jfrens/lexical_diversity) and processed 28,493,311 fanfiction chapters with minimum length 100 words. In total, 61,560,528,896 words were processed.

## Publication Time Estimation, Language, and MTLD Outliers

Chapter publication times are not directly accessible on the website, thus we made estimates using story and review metadata. We took the story publication time as the time for the first chapter. For subsequent chapters, we used the time of the first review as an estimate of its time of publication. To verify the accuracy of this estimate, we compared story publication time with first review time for first chapters, and found that the median time to review a first chapter was 3 days, and 42% of first chapters received their first review within 24 hours. Chapters with zero reviews were assigned publication times equal to the nearest known chapter times.

We obtained story languages from metadata available on Fanfiction.net. We verified the accuracy of this data using the Python library langdetect. Overall, the metadata matched with langdetect when finding English vs non-English for 99.5% of chapters, exceeding the 99% claimed accuracy of langdetect. MTLD varies drastically with language, and previous studies have utilized lemmatization with MTLD while working with non-English languages (Treffers-Daller, 2013). Our study included only English language texts, 25,266,230 out of 28,493,311 total chapters, and did not use lemmatization.

While most fanfiction chapters had MTLD between 50 and 150, a few texts had extremely low or high scores. We reviewed a sample of texts with MTLD below 5 and found that almost all of these low-scoring texts are non-narrative word repetitions. One author, in a final chapter, wrote "I LOVE YOU GUYS! HAVE ALL THE COOKIES! (::)(::) (::)(::) (::)(::)," continuing to repeat the cookie emoticon for dozens of lines. A sample of texts above MTLD 300 were mostly non-narrative, including number sequences, lists of random words, tables of contents, glossaries, and random typing. One author achieved the highest MTLD, over 2.5 million, with a chapter quoting a character counting from one to ten thousand. We eliminated 2,678 outlier chapters with MTLD below 5 or above 300 from the analysis. We also eliminated 22 chapters with erroneous data, and 427,662 chapters containing fewer than 100 words. The dataset used for our analysis of lexical diversity includes 53,185,524,320 words contained in 24,835,868 chapters of fanfiction from 5,906,217 stories. Chapter MTLD scores in this set were normally distributed around the mean of 97.35, with a standard deviation of 21.96.

### Mixed Linear Models

Mixed linear models are a class of regression models suited to testing longitudinal differences on a continuous dependent variable. In a mixed model, fixed effects represent independent variables of interest. Random effects typically account for individual differences, such as between students, and group differences, such as between classrooms. In our regression analyses, fixed effects were used to model our independent measures: cumulative reviews and time. Random effects were used to group data by user and by fandom. Fandom is an important confound to control, as Yin et al. (2017) found that the number of reviews exchanged varies by fandom, and we found that MTLD varies by fandom.

## Results

### Reviews and Incremental Change in Lexical Diversity

To test H1, we examined MTLD change between subsequent chapters written within a one-month window with respect to reviews. We calculated 19,709,160 MTLD differences for this analysis, with a mean increase of .019 (SD=20.69). We determined the number of reviews received by the author between chapter publications (Mean=4.51, SD=6.67). We used reviews and days as fixed effects and user as a random effect in our mixed linear model. The fixed effects were weakly correlated (r=0.30). The resulting coefficient for reviews (see Table 1) indicates that each additional review predicted a decrease in MTLD of 0.007, while the coefficient for days indicates that each day between chapters was associated an increased MTLD of 0.024. Cohen's F2 for both variables was <0.001, indicating the effect sizes were nominal. The results contradict H1, showing that increased numbers of reviews do not predict an immediate increase on the subsequently written chapter.

**Table 1: Fixed effect coefficients predicting MTLD differences between chapters. Columns included are coefficients (Coeff.), Standard Error (SE), and Cohen's $F^2$ ($F^2$).**

| Fixed Effect | Coeff. | SE | $F^2$ |
|---|---|---|---|
| Days Between | 0.024* | <0.001 | <0.001 |
| Reviews Betw. | -0.007* | <0.001 | <0.001 |

*$p < 0.001$

**Table 2: Fixed effect coefficients predicting MTLD based on maturation (days) and distributed mentoring abundance (previous reviews).**

| Fixed Effect | Coeff. | SE | $F^2$ |
|---|---|---|---|
| Days | 0.0032* | <0.001 | 0.004 |
| Prev. Reviews | 0.0018* | <0.001 | <0.001 |

*$p < 0.001$

### Age and Lexical Diversity

We examined the relationship between age and MTLD for English speaking Fanfiction.net authors who self-reported their ages (see Figure 1). The mean chapter MTLD increases from 93.6 at age 15 to 97.1 at age 19, and thereafter remains generally flat. To test H2, that lexical diversity increases during late adolescence, we analyzed the 1,608,824 chapters by 71,983 authors with estimated ages from 15.0 to 20.0 years old with a mixed linear regression. Age is the only fixed effect, while user and fandom are modeled as random effects. The significant (p<0.001) and positive coefficient of 1.66 indicates that MTLD substantially increases each year during late adolescence. Cohen's F2 is 0.007, indicating the effect size is small relative to variance. This result supports H2, replicating previous findings that show adolescence to be a significant period of lexical development (White, 2014).

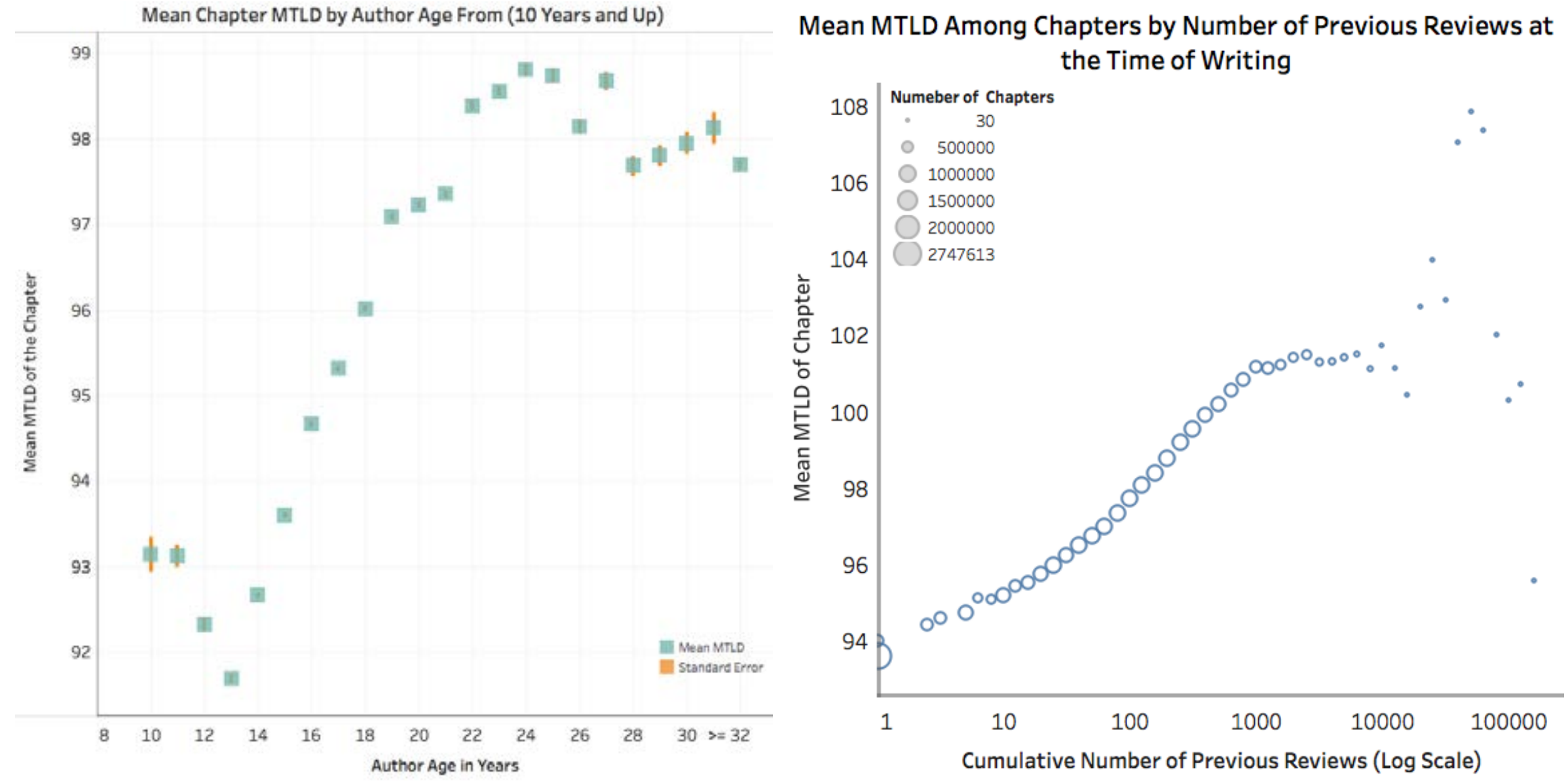

**Figure 1: Mean MTLD of chapters by author age in years. 154,451 authors are included. The rightmost point aggregates all chapters by authors aged 32 to 99.**

**Figure 2: Mean MTLD of chapters by cumulative number of reviews. The X axis is in log scale. The size of each circle indicates the number of chapters contained in each group.**

### Distributed Mentoring Abundance and Lexical Diversity

To operationalize the abundance of distributed mentoring, we counted, for each chapter in the English dataset (N=24,835,868), the cumulative number of previously received reviews by the same author. The median number of reviews was 59, with a right skew (Mean=420.38, SD=1741.70), and a maximum of 128,870 reviews. To visually examine the relationship between previous reviews and lexical diversity, we created logarithmic buckets of the chapters by the number of previously received reviews and computed the mean MTLD score among chapters in each bucket (see Figure 2). As shown in Figure 2, the mean lexical diversity (MTLD) increases with the number of previous reviews, from 93.22 when reviews are absent to 102.33 when over 10,000 reviews have been received by the author. We performed a mixed linear regression to test H3, that maturation will predict increased lexical diversity, and H4, that lexical diversity will increase as authors accumulate reviews. The dependent variable is MTLD, and fixed effects are previous reviews (abundance of distributed mentoring) and days since starting (maturation). The two fixed effects were weakly correlated (r=0.27). Fandom was modeled as a random effect. This analysis tracks MTLD changes during authors' first 50 chapters. 1,065,606 authors wrote at least two chapters, and 16,658,721 of their chapters are included in this analysis. Previous reviews and days each significantly predicted chapter lexical diversity (see Table 2). For each day of maturation, MTLD increased by .0032. For each review received on a previous work, MTLD increased by .0018. This analysis supports H3 and H4, showing that the abundance of distributed mentoring uniquely contributes to authors' development. Effect size as measured by Cohen's F2 was very small for days and almost negligible for previous reviews relative to MTLD variance.

## Limitations

Limitations and validity threats should be considered. First, there could be other causes for lexical diversity increase that correlate with distributed mentoring as operationalized by reviews. Second, our finding does not imply any causal relationship. Third, we do not know the degree to which stories were edited after being published. Furthermore, lexical diversity does not capture all aspects of narrative writing quality, nor does it represent all learning that occurs among fanfiction writers. This stems from broader issues: that no algorithm assesses text in the same way as a human evaluator, and no behavioral measure can peek into participants' minds to see what is learned.

## Discussion

We found that an abundance of distributed mentoring predicts increased lexical diversity among fanfiction chapters. This was robust when we accounted for maturation and fandom differences. Effect sizes (Cohen's $F^2$) were very small, indicating variance in MTLD is mostly predicted by factors other than distributed mentoring or maturation. It is unsurprising to find this high degree of noise in an automated learning measure. The results imply that reviews exchanged on Fanfiction.net shape authors' writing. Lexical diversity trends with narrative quality (Fergadiotis et al., 2015; Olinghouse & Wilson, 2013) and language ability (Mazgutova & Kormos, 2015; Treffers-Daller, 2013; White, 2014). Our findings contribute behavioral evidence in support of claims by young authors interviewed by Evans et al. (2017), that the community contributed to their development as writers. While reviews did not immediately increase lexical diversity on the subsequent chapter, the effect occurred over time as reviews accumulated. Receiving roughly 650 reviews predicted the same increase in lexical diversity as one year of maturation. This underscores the significance of informal writing communities in the lives of young writers and the importance of affordances for distributed mentoring in such communities.

Several implications follow from our analysis of the abundance of distributed mentoring, particularly for members of learning communities like Fanfiction.net. Participants in informal learning communities should be encouraged to embrace and interact with those who have not yet received feedback on their work. This type of community support can occur spontaneously, such as the "Review Revolution" on Fanfiction.net (Campbell et al., 2016), but the creation of affordances by community developers to facilitate review encouragement would likely yield a significant dividend for new writers. There are fundamental implications for stakeholders such as parents, teachers, designers and researchers. We need to recognize the role of fanfiction in shaping the development of today's connected youth. The type of feedback given through distributed mentoring has been discounted by researchers as shallow and therefore not valuable (Magnifico, Curwood, & Lammers, 2015). Our results contribute behavioral evidence to the growing number of ethnographic and qualitative studies demonstrating the importance of fanfiction for shaping the identities (Black, 2008), expression (Jenkins, 2006), and literacy (Chandler-Olcott & Mahar, 2002; Jamison, 2013) of young people. We should honor what young people are doing. Our findings support calls to acknowledge that this is a valid learning experience and incorporate it into formal education (Alvermann, 2008). Involved adults should encourage adolescent participation in informal writing communities so young writers can engage in and benefit from distributed mentoring.

This work opens areas for exploration in the study of connected learning in fanfiction communities. Evans et al.'s (2017) aspects of distributed mentoring provide a framework for exploration of reviews. Future work can extend ours by quantitatively examining different kinds of mentoring in the over 170 million reviews present on Fanfiction.net. We hypothesize that, given equal abundance of reviews, a greater diversity of review perspective and content will be associated with improved outcomes. Another potential direction comes from identifying and understanding roles that users take on within Fanfiction.net. As noted by Campbell et al. (2016), there is no overt distinction among users in their profile pages, especially age-based distinctions typical of offline settings, unless they elect to report this information. Thus, the context of Fanfiction.net provides teens and emerging adults with unique opportunities to assume mentorship roles. A network analysis is needed to review the roles that exist in the fanfiction community and how the roles of author and reviewer interact in the network and help to uncover design principles for incorporating distributed mentoring into other learning settings.

## Conclusion

Young adults, at an age critical to lexical development, represent the majority of Fanfiction.net users. This co-occurrence of development with fanfiction authorship, along with our found association between reviews and lexical diversity, underscore the importance of distributed mentoring in online writing communities for the growth of young authors. This study is the largest application of MTLD to a public corpus, as well as the first longitudinal analysis of writing at such a massive scale. Our findings support calls to promote reviewing behavior and incorporate fanfiction into formal learning. Work remains to further explore reader-reviewer relationships, examine aspects of distributed mentoring beyond sheer abundance, and assess how best to support mentorship in informal online learning communities.

## References

Alvermann, D. E. (2008). Why Bother Theorizing Adolescents' Online Literacies for Classroom Practice and Research? *Journal of Adolescent & Adult Literacy*, *52*(1), 8–19.

Black, R. W. (2008). *Adolescents and online fan fiction* (Vol. 23). Peter Lang.

Campbell, J., Aragon, C., Davis, K., Evans, S., Evans, A., & Randall, D. (2016, February). Thousands of positive reviews: Distributed mentoring in online fan communities. In *Proceedings of the 19th ACM Conference on Computer-Supported Cooperative Work & Social Computing* (pp. 691-704). ACM.

https://doi.org/10.1145/2818048.2819934

Chandler-Olcott, K., & Mahar, D. (2002). Adolescents' anime-inspired "fanfiction": An exploration of Multiliteracies. *Journal of Adolescent & Adult Literacy*, *46*(7), 556. https://doi.org/10.2307/40015457

Crossley, S. a., Salsbury, T., McNamara, D. S., & Jarvis, S. (2011). Predicting lexical proficiency in language learner texts using computational indices. *Language Testing*, *28*(4), 561–580. https://doi.org/10.1177/0265532210378031

Durán, P., Malvern, D., Richards, B., & Chipere, N. (2004). Developmental trends in lexical diversity. *Applied Linguistics*, *25*(2), 220–242+287. https://doi.org/10.1093/applin/25.2.220

Evans, S., Davis, K., Evans, A., Campbell, J. A., Randall, D. P., Yin, K., & Aragon, C. (2017). More Than Peer Production: Fanfiction Communities as Sites of Distributed Mentoring. *Cscw'17*. https://doi.org/10.1145/2998181.2998342

Fergadiotis, G., Wright, H. H., & Green, S. B. (2015). Psychometric Evaluation of lexical diversity indices: assessing length effects. *Journal of Speech, Language, and Hearing Research*, *58*(3), 840–852.

Hutchins, E. (1995). *Cognition in the Wild*. MIT press.

Jamison, A. (2013). *Fic: Why fanfiction is taking over the world*. BenBella Books, Inc.

Jenkins, H. (2006). Confronting the Challenges of Participatory Culture: Media Education for the 21 Century. *Chicago, IL: The MacArthur Foundation*.

Magnifico, A. M., Curwood, J. S., & Lammers, J. C. (2015). Words on the screen: Broadening analyses of interactions among fanfiction writers and reviewers. *Literacy*, *49*(3), 158–166. https://doi.org/10.1111/lit.12061

Mazgutova, D., & Kormos, J. (2015). Syntactic and lexical development in an intensive English for Academic Purposes programme. *Journal of Second Language Writing*, *29*, 3–15. https://doi.org/10.1016/j.jslw.2015.06.004

McCarthy, P. M., & Jarvis, S. (2010). MTLD, vocd-D, and HD-D: a validation study of sophisticated approaches to lexical diversity assessment. *Behavior Research Methods*, *42*(2), 381–92. https://doi.org/10.3758/BRM.42.2.381

McNamara, D. S., Crossley, S. a., & McCarthy, P. M. (2010). Linguistic Features of Writing Quality. *Written Communication*, *27*(1), 57–86. https://doi.org/10.1177/0741088309351547

Milli, S., & Bamman, D. (2016). Beyond Canonical Texts : A Computational Analysis of Fanfiction. *Proceedings of the 2016 Conference on Empirical Methods in Natural Language Processing (EMNLP-16)*, 2048–2053.

Olinghouse, N. G., & Wilson, J. (2013). The relationship between vocabulary and writing quality in three genres. *Reading and Writing*, *26*(1), 45–65. https://doi.org/10.1007/s11145-012-9392-5

Treffers-Daller, J. (2013). Measuring lexical diversity among L2 learners of French. *Vocabulary Knowledge: Human Ratings and Automated Measures*, *47*, 79.

White, R. H. (2014). Lexical richness in adolescent writing, insights from the classroom: An L1 vocabulary development study. https://doi.org/10.1007/s13398-014-0173-7.2

Yin, K., Aragon, C., Evans, S., & Davis, K. (2017). Where No One Has Gone Before: A Meta-Dataset of the World's Largest Fanfiction Repository. *CHI 2017 : ACM CHI Conference on Human Factors in Computing Systems*. https://doi.org/10.1145/3025453.3025720

Yu, G. (2010). Lexical diversity in writing and speaking task performances. *Applied Linguistics*, *31*(2), 236–259. https://doi.org/10.1093/applin/amp024